\documentstyle[12pt]{report} 
\def\rbuildrel#1\over#2{\mathrel{\mathop{#2}\limits_{#1}}}

\catcode`@=11
\def\underline#1{\relax\ifmmode\@@underline#1\else
        $\@@underline{\hbox{#1}}$\relax\fi}

\def\titlepage{\pagestyle{empty}\c@page=0
      \def\thefootnote{\fnsymbol{footnote}} }
\def\endtitlepage{\pagestyle{plain}\c@page=1
      \def\thefootnote{\arabic{footnote}} \c@footnote\z@ }
\catcode`@=12

\newskip\humongous \humongous=0pt plus 1000pt minus 1000pt
\def\caja{\mathsurround=0pt}
\def\eqalign#1{\,\vcenter{\openup1\jot \caja
        \ialign{\strut \hfil$\displaystyle{##}$&$
        \displaystyle{{}##}$\hfil\crcr#1\crcr}}\,}
\newif\ifdtup

\def\*{\hskip .06 cm}

\def\thebibliography#1{\section*{\ \markboth
 {REFERENCES}{REFERENCES}}\list
 {{\arabic{enumi}}.}
 {\settowidth\labelwidth{{#1}.}\leftmargin\labelwidth
 \advance\leftmargin\labelsep
 \usecounter{enumi}}
 \def\newblock{\hskip .11em plus .33em minus -.07em}
 \sloppy
 \sfcode`\.=1000\relax}

\def\thebibliographyp#1{\section*{\ \markboth
 {Chan \& Dill, $\quad$ Polymer Principles in Protein Structure
 and Stability}{Chan \& Dill, $\quad$ Polymer Principles in Protein Structure
 and Stability}}\list
 {\arabic{enumi}.}{\settowidth\labelwidth{#1.}\leftmargin\labelwidth
 \advance\leftmargin\labelsep
 \usecounter{enumi}}
 \def\newblock{\hskip .11em plus .33em minus -.07em}
 \sloppy
 \sfcode`\.=1000\relax}

\def\sqr#1#2{{\vcenter{\vbox{\hrule height.#2pt
\hbox{\vrule width.#2pt height#1pt \kern#1pt
\vrule width.#2pt}
\hrule height.#2pt}}}}

\begin{document}
%
\noindent
$\null$
\hfill April 15, 2003\\

\vskip 0.6in 

\begin{center}
{\Large\bf Simple Two-State Protein Folding Kinetics Requires}\\

\vskip 0.3cm

{\Large\bf Near-Levinthal Thermodynamic Cooperativity}\\

\vskip .5in 
{\bf H\"useyin K{\footnotesize{\bf{AYA}}}}
and
{\bf Hue Sun C{\footnotesize{\bf{HAN}}}}$^\dagger$\\
$\null$

Protein Engineering Network of Centres of Excellence (PENCE),\\
Department of Biochemistry, and
Department of Medical Genetics \& Microbiology,
Faculty of Medicine, University of Toronto,
Toronto, Ontario M5S 1A8, Canada\\

%

%

\end{center}

\vskip 1cm

\noindent
{\bf Running title:} Origins of Simple Two-State Protein Folding \\

\vskip 1cm

chevron rollover / non-Arrhenius kinetics /
G\=o models / contact order /

$\null$\\ 
%


\noindent 
$^\dagger$ Corresponding author.\\ 
E-mail address of Hue Sun C{\footnotesize{HAN}}: 
chan@arrhenius.med.toronto.edu\\
Tel: (416)978-2697; Fax: (416)978-8548\\
Mailing address: Department of Biochemistry, University of Toronto, 
Medical Sciences Building -- 5th Fl., 1 King's College Circle,
Toronto, Ontario M5S 1A8, Canada.

\vfill\eject 
%

\def\thefootnote{\fnsymbol{footnote}}

\noindent 
{\large\bf Abstract}\\

\vskip .2 in

\noindent 
Simple two-state folding kinetics of many small single-domain proteins 
are characterized by chevron plots with linear folding and unfolding 
arms consistent with an apparent two-state description of equilibrium 
thermodynamics. This phenomenon is hereby recognized as a nontrivial 
heteropolymer property capable of providing fundamental insight into protein 
energetics. Many current protein chain models, including common lattice and 
continuum G\=o models with explicit native biases, fail to reproduce this 
generic protein property. Here we show that simple two-state kinetics is
obtainable from models with a cooperative interplay between core burial and 
local conformational propensities or an extra strongly favorable energy 
for the native structure. These predictions suggest that intramolecular 
recognition in real two-state proteins is more specific than that 
envisioned by common G\=o-like constructs with pairwise additive energies.
The many-body interactions in the present kinetically two-state models 
lead to high thermodynamic cooperativity as measured by their van't Hoff 
to calorimetric enthalpy ratios, implying that the native and denatured
conformational populations are well separated in enthalpy by a high free 
energy barrier. It has been observed experimentally that deviations from 
Arrhenius behavior are often more severe for folding than for unfolding. This 
asymmetry may be rationalized by one of the present modeling scenarios
if the effective many-body cooperative interactions stablizing the native 
structure against unfolding is less dependent on temperature than the 
interactions that drive the folding kinetics.

\vfill\eject


\noindent
{\bf INTRODUCTION}

$\null$

A logical test of any general conception about the driving forces in 
protein folding is to ascertain whether polymer models incorporating
the given idea can predict generic behavior of real proteins.$^{1,2}$ 
In these considerations, self-contained polymer models with explicit chain 
representations$^{3}$ are of particular importance. Quite obviously,
the relationship between model energetics and conformational 
distribution can only be addressed in a physically plausible manner 
when chain connectivity and excluded volume are adequately taken into 
account.  Using this analytical framework, we found that even mundane 
protein properties such as calorimetric two-state cooperativity$^{4-6}$ and 
simple two-state folding/unfolding kinetics$^{7,8}$ are remarkable feats 
from a polymer standpoint. Simply put, it is nontrivial to construct 
heteropolymer models with commonly used model interaction schemes to 
reproduce such behavior. These include popular 2-, 3-, 20-letter models 
and traditional G\=o models (see below).$^{4-8}$ Generic protein properties 
thus present severe constraints on modeling. Hence, insight into real 
protein energetics can be gained by requiring self-contained polymer 
models to satisfy such constraints.
\\

Motivated by the proposed consistency principle$^{9}$ or principle of minimal 
frustration$^{10}$ for protein energetics, G\=o$^{11}$ and G\=o-like models 
(see, e.g., refs.~8, 12--16 and references therein) have long been used 
in protein folding investigations. These models postulate that only 
intrachain interactions found in the native (ground-state) conformation 
are favorable, all other possible intrachain interactions are assumed to be 
either neutral or unfavorable. Recently, this native-centric approach 
to modeling has often been justified as well by the discovery that folding 
rates of natural small single-domain proteins are well correlated with the 
contact order$^{17}$ of their native structures. How well do common G\=o models
mimic the generic properties of small single-domain proteins? For chain
models configured on two-dimensional square lattices, we found that 
even with their explicit native biases, the common G\=o interaction scheme 
falls far short of producing the type of calorimetric two-state cooperativity 
observed for many small proteins.$^{4}$ Three-dimensional G\=o-like 
lattice$^{5-7}$ and continnum$^{8}$ models are more proteinlike in this 
regard, as many of them may be considered calorimetrically cooperative 
if certain lattitude is allowed for empirical baseline subtractions.$^{5}$ 
\\

However, common G\=o-like schemes are apparently not capable of 
producing simple two-state folding/unfolding 
kinetics with linear chevron plots.$^{18}$ This we have recently demonstrated
in several examples,$^{7,8}$ including lattice and continuum (off-lattice) 
models as well as models with a rudimentary implicit-solvent 
treatment of desolvation barriers.$^{8,19}$ Thus, the inability of common 
G\=o-like constructs to predict simple two-state folding/unfolding kinetics 
is not an artifact restricted only to lattice G\=o models. Most likely, it
is a fundamental problem arising from the additive nature of common G\=o-like 
interaction schemes. Our results indicate strongly that such model 
interaction schemes --- on-lattice and otherwise --- afford insufficient 
cooperativity to capture real two-state protein energetics. Here we 
address this basic question by constructing and testing novel 
native-centric lattice models that go beyond common additive G\=o-like 
schemes, with intrachain potentials that can lead to simple two-state 
folding/unfolding kinetics. The ultimate goal of this line of inquiry, of 
which the present lattice exercise is only a first step, is to decipher 
the many-body cooperative interactions underlying the behavior of small 
single-domain proteins. 
\\

$\null$


\noindent
{\bf MODELING A COOPERATIVE INTERPLAY BETWEEN LOCAL}\\
{\bf CONFORMATIONAL PREFERENCE AND PROTEIN CORE}\\
{\bf FORMATION}

$\null$

We first explore several native-centric variants of a 55mer model 
(Figures~1--4). Their basic features are derived from the 
original model we put forth recently.$^{6}$ For the G\=o-like
constructs studied here, we retain contributions from the term 
disfavoring the initiation of left-handed helices (equation~1 of
ref.~6). Contributions from the 5-letter contact energies are retained 
for the native contacts in the ground-state conformation, 
whereas all nonnative contacts are assigned zero energy as in common 
G\=o models.  The resulting native-centric model has the energy function
$$
E= E_{\rm contact}^\prime + \gamma_{\rm lh}N_{\rm lh} \; ,
\eqno(1)
$$
where the prime superscript on the $E_{\rm contact}^\prime$ term indicates 
that the sum of pairwise 5-letter energies is restricted to native contacts,
and the second term on the right disfavors left-handed helices. Here we
use the same contact energies and $\gamma_{\rm lh}$ parameter as in ref.~6.
The ground-state energy of the present model equals $-36.1$. We refer to 
this as model (i).
\\

Next we consider a model that embodies the idea of a cooperative 
interplay between local conformational preference and the (nonlocal) 
packing of the protein core.$^{4,6,7}$ This is motivated by the observation 
that secondary structure formation in globular proteins is often context
dependent, and that short helices are often not stable in isolation
but are stable when packed in the core of a protein (see, e.g., ref.~20 
and references therein). Here we mimic this effect by assigning a
favorable energy ${\cal E}_{\rm coop}$ to each incidence of the
conformational situation described in Figure~1, leading to the
energy function
$$
E= E_{\rm contact}^\prime + \gamma_{\rm lh}N_{\rm lh} 
+ {\cal E}_{\rm coop} h_{\rm c} \; ,
\eqno(2)
$$
where $h_{\rm c}$ counts the incidences of a fully formed
native helix having the cooperative interactions defined by Figure~1,
and $0\le h_{\rm c}\le 4$ for the present 55mer model. This we
refer to as model (ii).
\\

To account for the possibility that optimal packing of the protein
core as a whole can impart significant thermodynamic stability
to the native structure, we consider yet another model with an extra
favorable energy assigned only to the native conformation. 
The energy function now becomes
$$
E= E_{\rm contact}^\prime + \gamma_{\rm lh}N_{\rm lh} 
+ {\cal E}_{\rm coop} h_{\rm c} + E_{\rm gs} \; ,
\eqno(3)
$$
where the augmented $E_{\rm gs}$ term takes a nonzero favorable value 
only when the chain is in its unique ground-state conformation.
We refer to this as model (iii).
We note that the ${\cal E}_{\rm coop}$ and $E_{\rm gs}$ terms introduced 
in equations~2 and 3 are many-body in nature. Many-body interactions have 
been investigated in the context of protein folding (see, e.g., refs.~6, 21, 
22). However, their relationship with linear chevron plots and simple 
two-state kinetics has not been much explored.
\\

Figure~2 shows that the many-body cooperative interactions introduced 
above enhance thermodynamic cooperativity. In calculating the heat 
capacities of these models, we made the simplifying assumption that 
the interactions are temperature independent, and set enthalpy equal to 
the model energy, as in our previous investigations.$^{5-8}$ 
For calorimetric two-state behavior, the van't Hoff to calorimetric 
enthalpy ratio $\Delta H_{\rm vH}/\Delta H_{\rm cal}$ has to be
close to one.$^{4-6}$ Now the $\Delta H_{\rm vH}/\Delta H_{\rm cal}$ ratio
($\kappa_2$ in ref.~5 without empirical baseline subtraction) equals 
$0.804$ for model (i) which does not contain 
many-body cooperative interactions. But it is considerably higher at $0.88$ 
and $0.91$, respectively, for models (ii) and (iii) with favorable values
of ${\cal E}_{\rm coop}$ and $E_{\rm gs}$ 
(Figure~2, upper panel).\footnote{
For every model considered in Figure~2, the $\kappa_2^{({\rm s})}$ (ref.~5) 
value for the $\Delta H_{\rm vH}/\Delta H_{\rm cal}$ ratio after empricial
baseline subtraction equals 1.0.}
This is not too surprising because the many-body cooperative interactions 
defined above tend to increase the energetic (enthalpic) separation between 
the ground-state and near ground-state conformations on one hand and 
the open unfolded conformations on the other, 
or disfavor conformations with intermediate energy (enthalpy),
or both.  Both of these effects would lead to higher calorimetric 
cooperativity.$^{4,5}$ The thermodynamic ramifications of these interactions 
are further explored in the lower panel of Figure~2, which covers a broad 
range of values for ${\cal E}_{\rm coop}$ and $E_{\rm gs}$. An additional
scenario in which an extra favorable energy for the ground-state is
augmented to model (i), viz.,
$$
E= E_{\rm contact}^\prime + \gamma_{\rm lh}N_{\rm lh} + E_{\rm gs} \;
\eqno(4)
$$
is also studied [curve (a)]. The trend observed in the lower panel of 
Figure~2 is that stronger many-body cooperative interactions of the type 
defined above generally lead to higher calorimetric cooperativity. However, 
there appears to be an upper limit on $\kappa_2$ ($\approx 0.96$) 
achievable by the helix-packing term alone [curve (b)], because at very 
high $-{\cal E}_{\rm coop}$ values it is possible that some intermediate 
non-ground-state conformations can become relatively stable (c.f. Figure~1).
\\

Figure~3 presents the chevron plots for the three models considered in the 
upper panel of Figure~2. To model folding and unfolding kinetics at different 
interaction strengths, an energetic scaling parameter $\epsilon$ is
introduced. At a given $\epsilon$, the effective energy of a conformation 
with energy $E$ (given by equations~1, 2 or 3) is equal to $-\epsilon E$;
and variation in $\epsilon/k_{\rm B}T$ (at constant $T$) serves
as a model denaturant concentration variation, as in ref.~7. Figure~3
shows that at sufficiently strong intrachain interaction (more negative
$\epsilon/k_{\rm B}T$), every folding arm of the three chevron plots
exhibits a rollover. This suggests that chevron rollover
is practically unavoidable in polymer models with physically plausible
interactions, because when intrachain interactions become generally
very favorable, kinetic trapping is bound to increase in importance.$^{7,8}$ 
However, native thermodynamic stability 
would be extremely high when the model parameter $\epsilon/k_{\rm B}T$ 
becomes extremely negative. Many such situations are not physically 
realizable in real proteins,$^{7}$ whose native stabilities even in
zero denaturant are often marginal. In this light, in comparing Figure~3
with experiments, the relevant question is whether there is a quasi-linear 
regime of the model chevron plots that is consistent with the two-state 
thermodynamics of the given model and covers a range of thermodynamic 
stability similar to that of real, simple two-state proteins.
\\

Pursuing this logic, we note that folding rollover occurs quite 
near to the transition midpoint for model (i), but the $\epsilon/k_{\rm B}T$
range of a quasi-linear regime is more extended for the two more 
cooperative models (ii) and (iii). The folding arms of models (ii) and 
(iii) are identical because, by construction, while the $E_{\rm gs}$ term 
in equation~3 slows unfolding, it does not affect the kinetics of folding.
For model (iii), we have used standard 
histogram techniques and extensive conformational sampling$^{7}$
at $\epsilon/k_{\rm B}T=-2.105$ to determine the dependence of
the free energy of unfolding $\Delta G_{\rm u}$ on $\epsilon/k_{\rm B}T$ 
(detailed data not shown). The resulting thermodynamic relation, which is 
essentially linear and has approximately the same transition midpoint as 
determined from the kinetic chevron plot, 
was applied to construct the dotted V-shape in 
Figure~3. The close agreement between the dotted V-shape 
and the simulated chevron plot for 
model (iii) from $\epsilon/k_{\rm B}T\approx -2.3$ to $-1.6$ implies that 
the folding/unfolding kinetics of model (iii) is consistent with a simple 
two-state description within this regime. The strongest intrachain 
interaction in the two-state regime is at $\epsilon/k_{\rm B}T\approx -2.31$, 
which corresponds to a native stability $\Delta G_{\rm u}\approx 10k_{\rm B}T$ 
for this particular model. It is clear from comparing the chevron plots
of models (ii) and (iii) that the linear regime can be readily extended
by increasing the magnitude of $E_{\rm gs}$ beyond that in model (iii).
But even as it stands, model (iii)'s behavior in Figure~3 should provide
a semi-quantitative rationalization for the simple two-state kinetics of 
many small single-domain proteins. Indeed, more than half of the 24 two-state 
proteins listed by Plaxco et al. (2000)$^{2}$ have native stabilities 
around 25$^\circ$C comparable to or lower than $10k_{\rm B}T$. For example, 
$\Delta G_{\rm u}=3.6k_{\rm B}T$ (2.1 kcal/mol) for CspB at 25$^\circ$C 
and pH 7.0 (ref.~23), and $\Delta G_{\rm u}=9.0k_{\rm B}T$ (5.3 kcal/mol)
for protein L at 22$^\circ$C and pH 7.0 (ref.~24). In our view, therefore,
simple two-state folding/unfolding kinetics emerges as a limiting-case 
phenomenon when the hypothetically high native stability at which chevron 
rollover would occur is not attainable by a small single-domain protein. 
Conversely, rollover becomes observable when a protein fails to achieve 
a sufficiently high thermodynamic cooperativity commensurate with its 
native stability.
\\

A comparison between the kinetic properties in Figure~3 and the 
thermodynamic properties in Figure~2 indicates that the thermodynamic
requirement of simple two-state behavior is stringent, allowing only
for a small adjustment from empirical baseline subtraction.$^{5}$
Apparently, a model has to be nearly or as cooperative as model (iii) 
or more to achieve a reasonable reproduction of simple two-state protein 
folding/unfolding kinetics. This suggests strongly that, in modeling
situations when the heat capacity contributions from bond vibrations are 
not considered (as in the present cases) and intrachain interaction 
energies are taken to be temperature-independent, a 
without-baseline-subtraction $\Delta H_{\rm vH}/\Delta H_{\rm cal}$ 
ratio of $\kappa_2 >0.9$ would likely be required for simple two-state 
kinetics [Figure~2, upper panel, model (iii)].
\\

We have also checked that folding kinetics is essentially single
exponential within the quasi-linear regime by confirming that the 
logarithmic distribution of folding first passage times is approximately 
linear.$^{7,8,25}$ In fact, extensive testing for ten values 
of $\epsilon/k_{\rm B}T$ between $-2.22$ and $-2.78$ covering 
the quasi-linear regime and beyond indicates that they are 
consistent with single exponential relaxation. Unfolding kinetics
is essentially single-exponential as well (detailed data not shown).
As in our previous study,$^{7}$ the onset of non-exponential folding 
relaxation at interaction strength $\epsilon/k_{\rm B}T\approx -2.9$ 
is concomitant to that of a drastic chevron rollover.
\\

Not surprisingly, Figure~4a shows that the free energy barrier
separating the native and denatured states is higher for a more 
cooperative model, consistent with its slower folding and 
unfolding rates at the transition midpoint (c.f. Figure~3). 
Figure~4b shows that the relation between
energy and the number of native contacts are approximately linear for
the two cooperative models. In this regard, the present exercise suggests 
that certain many-body interactions embodying a local-nonlocal cooperative 
interplay (${\cal E}_{\rm coop}=-1.0$) and an added ground-state 
stability ($E_{\rm gs}=-2.0$) in proteins can lead to remarkable 
improvements in kinetic cooperativity even when
the magnitudes of these terms are relatively small.
\\

$\null$

\vfill\eject


\noindent
{\bf MODELING A PARTIAL SEPARATION BETWEEN THE}\\
{\bf INTERACTIONS FOR THERMODYNAMIC STABILITY AND THE}\\
{\bf DRIVING FORCES FOR FOLDING KINETICS}

$\null$

Having established a plausible scenario for simple two-state protein 
folding/unfolding kinetics, we proceed to broaden our exploration
and to better delineate how various energetic components might contribute 
to this remarkable behavior. As a first step, we consider in this section a 
somewhat different class of models in which a local-nonlocal cooperative 
interplay is absent but the unique ground-state conformation is favored 
by an extra strong energy. The interaction scheme is a simplified version 
of equation~4 above, with the energy function
$$
E= E_{{\rm G}{\overline{\rm{o}}}}+E_{\rm gs} \; ,
\eqno(5)
$$
where $E_{{\rm G}{\overline{\rm{o}}}}$ is the usual lattice G\=o potential 
that assigns a favorable energy $\epsilon_0$ ($<0$) to every contact in 
the native structure of the model and assigns zero energy to all other 
(nonnative) contacts, and $E_{\rm gs}$ applies only to the ground-state
conformation, as in equations~3 and 4. To explore the effect of native 
topology, we study three cubic-lattice 27mer models with $E$ 
given by equation~5 for three different ground-state structures (Figures~5 
and 6).
\\

As discussed above, the $E_{\rm gs}$ term serves to increase native 
stability and enhance thermodynamic stability, leading to a reduced
unfolding rate. But it has no effect on the folding kinetics modeled 
by Monte Carlo dynamics with the Metropolis acceptance criterion.$^{26,27}$ 
Thus, the energetics described by equation~5 entails a partial separation 
between the interactions that drive the protein to fold kinetically (the 
pairwise contact energies $E_{{\rm G}{\overline{\rm{o}}}}$) and the
interactions that stabilize the ground-state structure 
($E_{{\rm G}{\overline{\rm{o}}}}$ and the many-body $E_{\rm gs}$). Since 
$E_{{\rm G}{\overline{\rm{o}}}}$ contributes partially to native stability, 
the role separation just described becomes a more predominant feature of 
the model when $E_{\rm gs}$ is large compare to $\epsilon_0$. A similar 
mechanism of partial separation between folding-kinetics and 
native-stabilizing interactions is also presumed by equation~3 
[model (iii)] and equation~4 above. Our interest in this scenario was 
partly motivated by experimental studies showing that mutants of a wildtype 
protein are much less likely to have a slower unfolding rate than to 
have a faster folding rate. For example, among the 41 mutants of Fyn SH3
domain studied by Northey et al.,$^{28}$ only 3 have slightly
reduced unfolding rates relative to that of the wildtype, whereas five
times as many (15) mutants have folding rates faster than that of the
wildtype. This means that interactions that accelerate folding do not
necessarily lead to higher native stability (only 4 mutants are
more stable than the wildtype), presumably because some mutants that 
fold fast do not pack well when folded$^{28}$ (c.f. discussion of
conformational strain by Ventura et al.$^{29}$). These observations
suggest that a partial separation of folding-kinetics and native-stabilizing
intraprotein interactions as envisioned by the $E_{\rm gs}$ term is 
physically plausible.
\\

Figure~5 shows that combining a pairwise G\=o potential with an $E_{\rm gs}$
term can also lead to simple two-state protein folding/unfolding kinetics,
although for these relatively short chains the $E_{\rm gs}/\epsilon_0$ 
ratio needed to achieve simple two-state behavior is large. Figure~5a
provides a series of unfolding chevron arms for different $E_{\rm gs}$ 
values, showing clearly that the quasi-linear regime of the chevron plot
can be extended by a more negative $E_{\rm gs}$. For all three models
considered in Figure~5 with $E_{\rm gs}/\epsilon_0=14$, approximately 
simple two-state behavior persists to $\epsilon/k_{\rm B}T\approx -1.15$ 
(c.f. simulated chevron plots and dotted V-shapes), corresponding to native 
stabilities $\Delta G_{\rm u}\approx 10 k_{\rm B}T$. As for model (iii)
above (Figure~3), for each model in Figure~5, we have verified that folding 
relaxation is essentially single exponential for $\epsilon/k_{\rm B}T>-1.6$ 
by obtaining linear logarithmic first passage time distributions for 
six $\epsilon/k_{\rm B}T$ values from $-1.67$ to $-0.91$. Unfolding
relaxation is also essentially single-exponential (detailed data not 
shown). Folding relaxation becomes non-exponential (for 
$\epsilon/k_{\rm B}T<-1.8$) only when native stability is much higher 
than that spanned by the simple two-state regime between 
$\epsilon/k_{\rm B}T\approx -1.15$ and $\epsilon/k_{\rm B}T\approx -0.723$. 
\\

Figure~6 compares chevron plots of the three cooperative 27mer models. 
The rank ordering of their folding rates is consistent with a correlation 
between slower folding rate and higher relative contact order (CO).$^{17}$
However, for these models, the dependence of folding rate on CO 
is weak. Near the onset of drastic chevron rollover and non-exponential 
folding relaxation ($\epsilon/k_{\rm B}T\approx -1.75$), the 27mer model 
with CO $=0.51$ folds only approximately 4 times slower than the 27mer 
model with CO $=0.28$. The dispersion in folding rate is even smaller
within the simple two-state regime. This is a far cry from the six orders 
of magnitude of variation in folding rates observed among real, small, 
single-domain proteins.$^{2}$ Recently, CO-dependent folding rates have been 
addressed using explicit-chain models with limited yet encouraging successes.
Using a G\=o-like potential for 18 small single-domain proteins,
Koga and Takada$^{31}$ obtained a correlation between CO and folding rate,
but the variation in rates covered only $\approx 1.5$ orders
of magnitude. More recently, Jewett et al.$^{30}$ conducted an 
extensive lattice 27mer simulation study using cooperative G\=o-like 
models with a nonlinear $E$--$Q$ relation. A correlation between CO
and folding rate was found but again the dispersion in folding rates 
spanned only $1$ to $1.5$ order of magnitude. While the mechanisms 
and energetics of CO-dependent folding remain to be better elucidated,$^{32}$ 
our more recent investigation shows that models with a local-nonlocal 
cooperative interplay similar to that in models (i) and (ii) 
(equations~2 and 3) above can lead to a relatively large dispersion in 
folding rates and a better correlation between CO and folding rate.$^{33,34}$
\\


\noindent
{\bf A RATIONALIZATION OF NON-ARRHENIUS PROTEIN}\\
{\bf FOLDING/UNFOLDING KINETICS}

$\null$

The physics embodied by the extra favorable ground-state energy 
$E_{\rm gs}$ in the models described by equations~3--5 above implies
that there is a fundamental asymmetry between folding and unfolding 
kinetics.$^{28}$ This led us to ask whether the same physical picture
may shed light on the significant difference in the degree of deviation 
from Arrhenius kinetics for folding versus unfolding that are often
observed in experiments.
Early measurements by Segawa and Sugihara$^{35}$ 
showed that the folding kinetics of hen egg-white lysozyme was 
significantly non-Arrhenius (logarithmic folding rate $\ln k_{\rm f}$ 
nonlinear in $1/T$) whereas the unfolding kinetics was essentially Arrhenius 
(logarithmic unfolding rate $\ln k_{\rm u}$ linear in $1/T$). Table~I 
summarizes more recent experimental data from the literature for several 
proteins with simple two-state folding/unfolding kinetics and whose 
temperature-dependent rates of both folding and unfolding have been measured 
directly.  For the proteins listed, the trend that folding is more 
non-Arrhenius than unfolding is quantified by reported activation heat 
capacities for folding $(\Delta C_p^\ddagger)_{\rm f}$ that are 
significantly larger in magnitude than the corresponding activation heat
capacity for unfolding $(\Delta C_p^\ddagger)_{\rm u}$. Table~I puts
the ``$(\Delta C_p^\ddagger)_{\rm f}/(\Delta C_p^\ddagger)_{\rm u}$'' ratio
in quotation marks because the common approach
of using temperature-independent activation heat capacities to analyze
folding/unfolding kinetics data may be problematic.$^{17}$ We note that
another potential source of the difficulty is that possible temperature 
dependencies of the heat capacities$^{39}$ associated with protein 
folding/unfolding transitions were not considered in such analyses. 
Nonetheless, as an empirical parameter, 
``$(\Delta C_p^\ddagger)_{\rm f}/(\Delta C_p^\ddagger)_{\rm u}$'' 
serves well to demonstrate that $\ln k_{\rm f}$ is often significantly 
more curvilinear in $1/T$ than $\ln k_{\rm u}$.
\\

This trend can be captured qualitatively in the present
modeling context if the solvent-mediated driving forces for folding 
kinetics and the many-body native-stabilizing interactions are taken
to have different temperature dependencies. This is a physically plausible 
assumption because some intraprotein solvent-mediated forces such as 
the hydrophobic effect are known to be sensitive to the sizes and shapes 
of the interacting groups.$^{40-42}$ Here we use a collection of 
cooperative 27mer models in Figure~5a with different values of
$E_{\rm gs}$ to expound the principles involved.
Temperature-dependent interactions are now introduced by letting the pairwise 
contact energy $\epsilon_0$ to vary with temperature in a hydrophobic-like 
manner while leaving $E_{\rm gs}$ temperature-independent.
\\

The following schematic analysis of $T$-dependent folding and unfolding
rates (Figure~7) is similar to that introduced by Chan$^{26}$ and Chan 
and Dill.$^{27}$ However, the present focus on folding-unfolding asymmetry 
in the context of three-dimensional protein chain models was not addressed 
in these earlier studies of short-chain two-dimensional models.$^{26,27}$ 
The first step in the 
present analysis is to obtain from Figure~5a the logarithmic folding and 
unfolding rates $\ln k_{\rm f}$ and $\ln k_{\rm u}$ [which are taken to be 
their respective $-\ln ({\rm MFPT})$] as functions of $\epsilon_0$ and 
$E_{\rm gs}$.  Since the effective energy is given by $-\epsilon E$
in Figure~5 (see above) where $E$ is given by equation~5 with $\epsilon_0$
set to $-1$, each $(\epsilon,E_{\rm gs})$-dependent datapoint in Figure~5a 
may be regarded as the folding or unfolding rate for the energy function 
$E$ itself with an $\epsilon_0$ value equals to that of $\epsilon$ and an 
$E_{\rm gs}$ value equals $\epsilon$ times the $E_{\rm gs}$ for the given 
unfolding chevron arm. We note that within the quasi-linear 
single-exponential regime,
$$
\ln k_{\rm f} = \alpha_{\rm f} + 
                \beta_{\rm f}{\frac {\epsilon_0}{k_{\rm B}T}} \;
\eqno(6)
$$
holds approximately for constant $\alpha_{\rm f}$ and $\beta_{\rm f}$,
because folding kinetics is independent of $E_{\rm gs}$.
A least-square fit yields $\beta_{\rm f}=-15.4$. For unfolding within
the quasi-linear single-exponential regime, the approximate linear relation
$$
\ln k_{\rm u} = \alpha_{\rm u} 
                + \beta_{\rm u}{\frac {\epsilon_0}{k_{\rm B}T}} 
                + \beta_{\rm u}^\prime {\frac {E_{\rm gs}}{k_{\rm B}T}} \;
\eqno(7)
$$
is expected, where $\alpha_{\rm u}$, $\beta_{\rm u}$, and 
$\beta_{\rm u}^\prime$ are constants. Extensive analyses indicate
that $\beta_{\rm u}\approx 4.0$ and $\beta_{\rm u}^\prime\approx 1.0$. 
We use $\beta_{\rm u}=3.9$, $\beta_{\rm u}^\prime=1$ below. Figure~7a 
shows that these values fit the simulated unfolding rates of the more 
cooperative models extremely well.
\\

Next, an hypothetical temperature-dependent $\epsilon_0$ $=\epsilon_0(T)$ 
is introduced in Figure~7b. This functional form for
$\epsilon_0/k_{\rm B}T$ (solid curve, left scale) was motivated by the 
temperature dependence of hydrophobic effects$^{39}$ and is similar to 
that explored in refs.~26 and 27. It follows that the temperature-dependent 
folding rate in the quasi-linear single-exponential regime is now given by
equation~6 above with $\epsilon_0$ $\rightarrow$ $\epsilon_0(T)$
provided by Figure~7b, viz.,
$$
\ln k_{\rm f} (T) = \alpha_{\rm f} + 
                \beta_{\rm f}{\frac {\epsilon_0(T)}{k_{\rm B}T}} \; .
\eqno(8)
$$
Similarly, the temperature-dependent unfolding rate in the quasi-linear 
single-exponential regime of an effective $E_{\rm gs}=-14$ cooperative 
model is given by equation~7 above with $\epsilon_0$ $\rightarrow$ 
$\epsilon_0(T)$ provided by Figure~7b while $E_{\rm gs}$ remains 
temperature independent:
$$
\ln k_{\rm u} (T) = \alpha_{\rm u} 
                    + \beta_{\rm u}{\frac {\epsilon_0(T)}{k_{\rm B}T}} 
                    + {\frac {E_{\rm gs}}{k_{\rm B}T}} \; ,
\eqno(9)
$$
where the $E_{\rm gs}/k_{\rm B}T$ term is set equal to $-14$ for a 
reference temperature ($T^*$) in Figure~7c at which 
$\epsilon_0/k_{\rm B}T=-1$. Hence $E_{\rm gs}/k_{\rm B}T=-14(T^*/T)$
is linear in $1/T$. These temperature-dependent folding and unfolding
rates are plotted in the upper part of Figure~7c. It is clear that
folding is significantly more non-Arrhenius than unfolding because 
the only source of non-Arrhenius behavior in the present formulation of
the model is $\epsilon_0(T)$, and $\ln k_{\rm f}$ depends more strongly 
on $\epsilon_0(T)$ 
($\beta_{\rm f}=-15.4$) than $\ln k_{\rm u}$ ($\beta_{\rm u}=3.9$).
\\

One missing physical ingredient in the consideration thus far is that
intrinsic conformational transition rates should accelerate at
higher temperature. This is not taken into account if physical time is 
simply identified with number of attempted Monte Carlo moves, as in
the analysis above. The issue has been identified and addressed in 
some detail in refs.~26 and 27. As in these
references, Figure~7b introduces an adjustment factor (dotted line, 
right scale) to better mimic physical time. Here $A(T)$ is the
temperature-dependent time needed for a given kinetic process 
and $A_0$ is a reference time. Thus the hypothetical $-\ln [A(T)/A_0]$ 
function in Figure~7b stipulates
that the intrinsic logarithmic rate [$-\ln A(T)$] is higher at
higher temperatures (i.e., the Monte Carlo clock should run faster 
at higher $T$). This adjustment factor is readily incorporated$^{26,27}$
by setting
$$
\eqalign{
\ln ({\rm folding\ rate}) & = \ln k_{\rm f} (T) - \ln[A(T)/A_0]\; , \cr
\ln ({\rm unfolding\ rate}) & = \ln k_{\rm u} (T) - \ln[A(T)/A_0] \; , \cr}
\eqno(10)
$$
where $\ln k_{\rm f}$ and $\ln k_{\rm u}$ on the right hand side are the 
expressions given respectively by equations~8 and 9. Macroscopically,
this amounts to introducing an additional enthalpic contribution$^{24,26}$
to the free energy barrier of protein folding. The lower part of
Figure~7c shows that incorporating a $-\ln[A(T)/A_0]$ term can
lead to a more realistic description of temperature-dependent
protein folding and unfolding rates (c.f. experimental data in Figure~7d;
see also Figure~3 of ref.~23, Figure~1C of ref.~24, Figure~3 of ref.~36, 
and Figure~5 of ref.~38). 
\\

$\null$


\noindent
{\bf DISCUSSION: A NEAR-LEVINTHAL SCENARIO FOR SIMPLE}\\
{\bf TWO-STATE PROTEIN FOLDING/UNFOLDING KINETICS}

$\null$

The results of the present investigation suggest strongly that the physical
interactions underlying the simple two-state folding/unfolding kinetics 
of small single-domain proteins should involve many-body effects beyond 
that stipulated by common additive G\=o models, even though the 
physico-chemical origins of these effects remain to be elucidated. 
Therefore, with regard to protein thermodynamic and kinetic cooperativity, 
common G\=o models with pairwise additive contact energies are not ideal. 
Apparently, the type of many-body interactions that are conducive to 
simple two-state kinetics also lead to higher thermodynamic cooperativity, 
and entail a partial separation between folding-kinetics and 
native-stabilizing interactions. Historically, an impetus to formulate 
the Levinthal paradox might have been the discovery in the late 1960s that 
some proteins were calorimetrically two-state$^{43}$ (see ref.~4 and 
references therein). Naturally, an extreme interpretation of the 
$\Delta H_{\rm vH}/\Delta H_{\rm cal}=1$ property would imply that only 
two enthalpy levels exist (native and denatured), and thus the landscape 
should resemble a golf course (Figure~8a). However, a golf-course landscape 
dictates that folding would be exceedingly slow, but the folding of real 
proteins is relatively fast. To address the ``why is folding fast'' 
question, recent theoretical discussions emphasize the funnel-like nature 
of the protein folding energy landscape as a solution to the Levinthal 
paradox$^{44,45}$ (Figure~8b); and common G\=o potentials are often used 
to model a relatively smooth funnel-like energy landscape.$^{12}$ We found 
that calorimetric two-state cooperativity can be consistent with the 
funnel-like landscapes of three-dimensional G\=o models, provided some 
lattitude is allowed for empirical baseline subtractions. This is because 
in these native-centric models, the conformational populations with 
intermediate energies (enthalpies) --- though not zero --- 
are relatively low.$^{5-8}$
\\

However, as discussed above, common G\=o models are insufficient for
simple two-state protein folding kinetics.$^{7,8}$ Because kinetic traps 
are still significant in these constructs under native conditions, the 
chevron plots they predicted have severe rollovers.$^{46}$ 
Another shortcoming of common G\=o models is that their 
predicted folding rates are often too fast compared to that of real 
proteins.$^{8,14}$ In one example, it is at least four orders of magnitude 
faster.$^{8}$ So, in a sense, in the context of recent G\=o modeling
efforts, the critical question has shifted from ``why is folding fast''
to ``why is folding slow.'' The present study concludes that a thermodynamic 
cooperativity higher than that afforded by common additive G\=o models 
is necessary for simple two-state kinetics (Figures~2 and 3). This scenario 
also offers a clue to the ``why is folding slow'' question. For real, small, 
single-domain proteins, it appears that the key to avoiding kinetic traps 
and chevron rollover is to have only weakly favorable intrachain interactions
during the folding process (gentle upper slopes of the funnel in Figure~8c)
until a significant fraction of the chain is native-like and ready to come 
together to form a large number of native contacts at once, at which point 
strong cooperative many-body effects kick in to stabilize the structure
(steep lower slopes of the funnel in Figure~8c). This idea is implemented in 
the cooperative models studied here. Indeed, in Figures~3 and 5, the 
quasi-linear folding regimes of the cooperative models are in
the weakly-interacting unfolding regimes (small $-\epsilon/k_{\rm B}T$) of 
the corresponding additive G\=o models.  Thus, for folding of the 
cooperative models, the energetic bias is not very strong during most of 
the conformational search. This feature serves to diminish the effects 
of kinetic traps (shallow minima on the gentle upper slopes of Figure~8c), 
because the depths of kinetic traps in heteropolymers are often correlated 
with the overall energetic bias towards the native structure.$^{27}$ As a 
result, folding is faster in the cooperative models relative to 
other heteropolymer models with deep kinetic traps. But at the same time, the 
very feature of a weakened energetic bias towards the native structure
during most of the conformational search also leads to slower folding in 
comparison with common G\=o models, because the latter have stronger 
native biases during the corresponding kinetic process. Nonetheless, the 
reduction in folding rate relative to common G\=o models does not make the 
cooperative models less proteinlike, because even a small bias is sufficient 
to circumvent the Levinthal paradox, as all of our models fold. In this way,
the cooperative model scenario provides a physical plausible answer to 
the ``why is folding slow'' question posed above. In fact, for 
real proteins, folding rates may be further reduced by the inevitable 
presence of favorable nonnative interactions, which are not taken into 
account by the native-centric cooperative models here. Anti-cooperativity of 
certain hydrophobic interactions$^{47}$ may also play a role in discouraging 
premature chain collapse. In this scenario, the energy landscape of a simple
two-state protein is still funnel-like but with a narrow bottleneck 
(Figure~8c). The resulting highly bimodal distribution of energy 
and high thermodynamic cooperativity thus approach (though never equal) 
that of an hypothetical Levinthal golf course.\footnote{
We emphasize that the relatively smooth funnel drawings in Figure~8 should be
viewed only as pictorial devices for underscoring the smoothness of the 
energy landscapes of native-centric models {\it relative} to that of models 
with deeper kinetic traps. Even for G\=o and G\=o-like models, energy
landscapes cannot be completely smooth because of repulsive interactions 
(including excluded volume effects) and other microscopic energy barriers 
due to bond rotations and solvation, for example (refs.~8, 26, 27; 
c.f. equation~10).
}
\\

Part of the present view is similar to that of Jewett et al.,$^{30}$
who recently introduced a native-centric model in which the energy $E$ 
of a conformation --- unlike that in common G\=o models --- does not 
decrease linearly with the number of native contacts. (Conformations 
with more negative $E$'s are more favorable.) In their model, the rate of 
decrease in $E$ with increasing fractional number of native contact $Q$ 
becomes progressively higher as the native structure is approached (when
$Q\rightarrow 1$). This means that the energetic bias towards the native 
structure is not strong during the initial stages of folding when there 
are few native contacts (small $Q$), but becomes stronger when the native 
structure is approached during the final stages of folding ($Q\rightarrow 1$). 
Insofar as this general trend is concerned, the physical picture 
of cooperative folding discussed above (especially the model defined 
by equation~5) is very much similar to that of Jewett et al.
Nevertheless, although both the model of Jewett et al. and the present 
cooperative models have high degrees of thermodynamic cooperativity,
their underlying mechanisms are not identical. More recent investigations 
indicate that detailed kinetic features, such as the correlation between 
CO and folding rate, do depend significantly on how thermodynamic 
cooperativity is achieved microscopically.$^{33,34}$ Since the lattice model
of Jewett et al.  was inspired by the more general topomer search model 
of folding,$^{48}$ it would be extremely interesting to compare in future 
investigations the relationships between the topomer search model and the 
several different scenarios of thermodynamic cooperativity explored here.
\\

As we have emphasized,$^{4,5}$ the experimental calorimetric two-state 
criterion, which has proven useful for evaluating protein chain 
models,$^{4-8,33,34,49-52}$ does not imply that there are only two 
infinitely sharp energy (or enthalpy) levels. In other words, 
the calorimetric two-state criterion does not preclude the existence of 
``partially unfolded'' conformations with energies
intermediate between the energy distribution peaks under strongly 
folding and strongly denaturing conditions (c.f. Figure~16 of ref.~4, 
Figure~10 of ref.~5, Figure~1 of ref.~7, and Figures~7--9 of ref.~8).
These theoretical findings are consistent with native-state hydrogen 
exchange experiments.$^{53}$ Conformational populations with intermediate 
energies in several calorimetrically cooperative models tested thus
far (see figure references above) share some similarities with that in 
calorimetrically non-cooperative constructs such as certain HP models$^{4}$ 
and a 15mer 20-letter sidechain model.$^{5,54,55}$ However, the critical 
difference between calorimetric cooperative and non-cooperative models is 
that, at the transition midpoint, conformational populations with 
intermediate energies are not significant for calorimetrically cooperative 
models but are significant for calorimetrically non-cooperative models. 
This difference is well characterized by the 
$\Delta H_{\rm vH}/\Delta H_{\rm cal}$ ratio.$^{4,5,49-52}$
Our approach of evaluating chain models by experimental cooperativity 
criteria is designed to address what kind of elementary intrachain 
interactions may be needed to produce the generic cooperative features 
of proteins, while taking into account as much as possible that proteins 
are polymers and therefore chain connectivity and excluded volume 
are severe constraints. In this respect, investigations using self-contained 
polymer models such as that conducted here are fundamentally more 
informative than those that use postulated free energy profiles in the 
absence of explicit chain representations (see, e.g., ref.~56). 
\\

In summary, our results suggest that intramolecular recognition in 
real two-state proteins is highly specific. As well, the role of many-body 
interactions in providing a larger average energetic difference 
between ``native'' and ``denatured'' conformations than that
afforded by common pairwise additive interaction schemes have 
potentially important implications for the discrimination of decoys 
in protein structure prediction.$^{57}$ In principle, the many-body 
interactions proposed in the present work can be characterized 
quantitatively through careful experiments and extensive atomic
simulations. How sidechain packing, sidechain/mainchain correlation,$^{58}$ 
and interactions such as hydrogen bonding contribute to this mechanism 
remains to be investigated. To help address these questions, the 
ramifications of the different scenarios explored in this work need to
be first delineated in greater detail.$^{33,34}$ Lattice model studies 
of Klimov and Thirumalai showed that sidechain degrees of freedom increase 
the sharpness of the thermodynamic folding/unfolding transition relative to 
that of the corresponding (mainchain) model with no sidechains.$^{54}$ 
However, their short-chain 20-letter sidechain models configured on 
three-dimensional simple cubic lattices do not appear to be calorimetrically 
cooperative. The $\Delta H_{\rm vH}/\Delta H_{\rm cal}$ ratio of one of 
the sidechain sequences studied in refs.~54 and 55 is equal to
$\kappa_2=0.38$ without baseline subtraction, and increases only
to $\kappa_2^{({\rm s})}=0.54$ after reasonable baseline subtractions.$^{5}$
These values are far from the $\Delta H_{\rm vH}/\Delta H_{\rm cal}\approx 1$
required for calorimetrically cooperative behavior. Kinetically,
even a G\=o-like version of their sidechain model exhibits a severe 
chevron rollover (Figure~3 of ref.~59). These results imply that
while sidechain contributions are expected to be important for protein
cooperativities,$^{54}$ their role has yet to be better elucidated.
\\

Finally, while the present study focuses on the 
behavior of small single-domain proteins, we hasten to emphasize
that not all proteins have simple two-state folding/unfolding kinetics.
Hence, the high cooperativity requirement deduced in the above analysis
may not apply to other proteins. In fact, one distinct advantage
of the energy landscape perspective and self-contained polymer modeling 
is their ability to cover a wide spectrum of possible protein behavior
under a unified physical framework. For instance, although 
common additive G\=o models are insufficient for simple two-state kinetics,
they are useful for understanding real proteins with chevron 
rollovers.$^{7,8,46,60}$ And even calorimetrically non-cooperative 
models (see discussion in ref.~5) may prove to be helpful in rationalizing 
downhill protein folding$^{61}$ as well.
\\


$\null$


\noindent
{\Large Acknowledgments}

We thank Yawen Bai, Robert L. Baldwin, Alan Davidson, Julie Forman-Kay, 
Michael Levitt, Kevin Plaxco, Steve Plotkin, Boris Steipe, 
and Dev Thirumalai for helpful 
discussions, and Kevin Plaxco for kindly providing ref.~30 before 
publication.  This work was partially supported by the Canadian Institutes of 
Health Research (CIHR grant no. MOP-15323), a Premier's Research 
Excellence Award from the Province of Ontario, and the Ontario Centre 
for Genomic Computing at the Hospital for Sick Children in Toronto.
H. S. C. is a Canada Research Chair in Biochemistry.


\par\vfill\eject

\noindent
{\large\bf References}

\kern -1.5cm

\vfill\eject

\noindent
{\large \bf Table I.} Deviations from Arrhenius behavior in 
protein folding and unfolding.
\vskip .2 in

\begin{center}
\begin{tabular}{|l|c|}
\hline
Protein name & 
``$(\Delta C_p^\ddagger)_{\rm f}/(\Delta C_p^\ddagger)_{\rm u}$''\\
\hline
T4 lysozyme mutant$^{\rm a}$ & $-2.99$\\
CI2$^{\rm b}$ & $-2.95$\\
CspB$^{\rm c}$ & $-9.0$\\
Protein L$^{\rm d}$ & $-1.68$\\
NTL9$^{\rm e}$ & $-1.33$\\ 
\hline
\end{tabular}
\end{center}
\vskip .15 in

\noindent
$^{\rm a}$ Chen {\it et al}. (ref.~36)\\
$^{\rm b}$ Jackson \& Fersht (ref.~37)\\ 
$^{\rm c}$ Schindler \& Schmid (ref.~23)\\
$^{\rm d}$ Scalley \& Baker (ref.~24)\\
$^{\rm e}$ Kuhlman {\it et al}. (ref.~38)\\

\vfill\eject

$\null$
\vskip -1.4cm

\noindent
{\large\bf Figure Captions}\\

\noindent
{\bf Figure 1.} $\quad$
Modeling a cooperative interplay between local conformational preference 
and protein core packing in a 55mer native-centric four-helix-bundle model. 
Certain native helices are shown as thick solid lines for illustrative 
purposes. Their sequence positions are 
indicated by the thin dotted lines depicting the rest of the full native 
structure. Given a native helix is completely formed [the front right helix 
appearing in both (a) and (b) in this example], a favorable 
cooperative energy ${\cal E}_{\rm coop}$ ($<0$) is assigned if either (a)
4 or more of the 6 native contacts (thick dotted lines) between the 
given fully formed native helix and each of the two chain segments for 
the two flanking native helices are present, or (b) at least 2 of the 3 
``diagonal'' native contacts are present between the given fully formed 
native helix and the chain segment for the diagonally neighboring native 
helix, or both [(a) and (b)]. Thus, condition (a) requires at least 8 
inter-helix nearest-neighbor native contacts, whereas condition (b) 
requires at least 2 inter-helix next-nearest-neighbor native contacts 
on the simple cubic lattice. It follows that the maximum total contribution 
from these cooperative energies is $4{\cal E}_{\rm coop}$ when all four 
native helices are completely formed and correctly packed against
one another.
\\

\noindent
{\bf Figure 2.} $\quad$
The overall thermodynamic cooperativity of a model protein is boosted by 
the many-body interactions described in Figure~1 and in the text.
{\bf Upper panel:} Heat capacity as a function of temperature
for {\bf (i)} the pairwise additive native-centric 55mer model (equation~1),
{\bf (ii)} a model having the interactions in (i) plus the cooperative 
interaction scheme in Figure~1 with ${\cal E}_{\rm coop}=-1.0$ (equation~2), 
and {\bf (iii)} a model with the interactions in (ii) plus an extra 
favorable energy of $E_{\rm gs}=-2.0$ for the ground-state conformation
(equation~3). The heat capacity scans were obtained using Monte Carlo 
histogram techniques based on conformational sampling around 
each model's transition midpoint.$^{5-7}$ The inset (from ref.~7) shows 
the 55mer model ground-state structure with (nominally) hydrophobic and polar 
residues depicted respectively as filled and open circles.
{\bf Lower panel:} Van't Hoff to calorimetric enthalpy 
$\Delta H_{\rm vH}/\Delta H_{\rm cal}$ ratios are given 
by $\kappa_2$ defined in 
ref.~5 (without empirical baseline subtractions) for three classes of 55mer 
models whose interaction schemes are parametrized by an ${\cal E}_{\rm coop}$ 
variable. {\bf (a)} As in (i) above plus an extra favorable energy of 
${\cal E}_{\rm coop}$ for the ground-state conformation (equation~4). 
{\bf (b)} As in (ii) above but, instead of fixing ${\cal E}_{\rm coop}=-1.0$, 
a variable ${\cal E}_{\rm coop}$ is used for the helix packing contribution 
defined by Figure~1 and equation~2. {\bf (c)} As in (b) plus an extra 
favorable energy of $E_{\rm gs}={\cal E}_{\rm coop}$ for the ground-state 
conformation (equation~3).
\\

\noindent
{\bf Figure 3.} $\quad$
Chevron plots for the 55mer models (i), (ii), and (iii) in Figure~2 are 
provided by the negative natural logarithm of mean first passage time (MFPT)
as functions of interaction strength $\epsilon/k_{\rm B}T$. The present
Monte Carlo (MC) dynamics simulations use the same general procedure as that 
in ref.~7, now with a move set consisting of end flips (4.7\%), corner 
flips (58.3\%), crankshafts (27\%) and rigid rotations (10\%).
Folding (filled symbols) starts from a randomly generated 
conformation. First passage time (FPT) for folding is the number of 
attempted MC moves needed to reach the ground-state conformation. Unfolding 
(open symbols) starts from the ground-state conformation. Here unfolding FPT
is the number of attempted MC moves needed to reach a conformation with
fewer than 7 native contacts. Each plotted MFPT is averaged from 1,000 
trajectories. Solid and dashed curves through the data points are mere 
guides for the eye. The dotted V-shape for model (iii) is an hypothetical 
simple two-state chevron plot consistent with the 
$\epsilon/k_{\rm B}T$ dependence of thermodynamic stability as given by 
the free energy difference $\Delta G_{\rm u}$ between the ground state 
and the unfolded conformational ensemble with $< 7$ native contacts.
\\

\noindent
{\bf Figure 4.} $\quad$
Distribution of native contacts in the 55mer models. The total (maximum) 
number of spatial nearest-neighbor native contacts in the ground-state
conformation is 60 (diagonal contacts in Figure~1b are not included in
this accounting). $Q$ is the fractional number of native contacts for a 
conformation, defined as the number of native contacts it contains divided 
by the maximum. (a) The free energy profiles of the models (i), (ii) and 
(iii) defined in Figure~2 are given by the negative logarithmic distributions 
of $Q$.  The profiles shown are for $\epsilon/k_{\rm B}T=$ $-2.34$, $-2.11$, 
and $-2.0$ respectively for (i), (ii) and (iii), near each model's 
transition midpoint, and were obtained by standard MC histogram 
techniques.$^{5-7}$ 
(b) Correlation between energy $E$ and $Q$ for models (ii) and (iii).
Dots indicate the existence of conformations with the given $E$ and $Q$
values. The open diamond and square mark the ground-state energies of
$-40.1$ and $-42.1$ for models (ii) and (iii) respectively.
\\

\noindent
{\bf Figure 5.} $\quad$
Chevron plots for three different 27mer G\=o models and for their 
corresponding models with an extra favorable energy $E_{\rm gs}$ assigned 
to the native (ground-state) structure (as shown), 
with $\epsilon_0=-1$ (equation~5). Folding MFPT is independent 
of $E_{\rm gs}$, and $\epsilon$ has the same meaning as in Figure~3.
Here each MFPT is averaged from 500 trajectories. Folding (filled 
symbols) and unfolding (open symbols) simulations were performed as for 
Figure~3 except only local chain moves were used for the present 27mer 
MC dynamics simulation (no rigid rotations), and unfolding FPT 
is now defined by the time needed to reach a conformation with fewer 
than 4 native contacts from a given starting ground-state conformation 
with 28 native contacts. Solid curves are mere guides for the eye. 
(a) Unfolding chevron arms for eight models with different degrees of 
cooperativity are shown; from top to bottom, $E_{\rm gs}=$ $0$ (common 
additive G\=o model), $E_{\rm gs}=$ $-2$, $-4$, $-6$, $-8$, $-10$, $-12$, 
and $-14$. (b) and (c) Unfolding 
chevron arms for the common additive G\=o model ($E_{\rm gs}=0$, 
upper curves) are compared with the unfolding arms for 
$E_{\rm gs}=-14$ (lower quasi-linear curves). The dotted V-shapes 
are hypothetical simple two-state chevron plots consistent with the 
$\epsilon/k_{\rm B}T$ dependence of $\Delta G_{\rm u}$ between the ground 
state and the unfolded conformational ensemble with $< 4$ native 
contacts of a given model; $\Delta G_{\rm u}$ values are determined
by standard histogram techniques$^{5-7}$ based on conformational sampling
at $\epsilon/k_{\rm B}T=-0.935$, $-0.917$ and $-0.917$ for
(a), (b) and (c) respectively (detailed data not shown).
\\

\noindent
{\bf Figure 6.} $\quad$
Comparing chevron plots for the three different models in Figure~5 
with $E_{\rm gs}=-14$ shows that a model with lower native 
contact order (CO, as defined in ref.~17) tends to fold slightly faster. 
Here CO $=$ $0.28$, $0.40$, and $0.51$ for models (a), (b) and (c) 
respectively.
\\

\noindent
{\bf Figure 7.} $\quad$
Rationalizing non-Arrhenius protein folding and unfolding kinetics.
(a) A linear fit of the logarithmic unfolding rates (vertical axis) of 
the 27mer models in Figure~5a with $-E_{\rm gs}\ge 6$ to the 
expression shown (horizontal axis). For the data points plotted,
the correlation coefficient $r=0.998$. 
(b) Solid curve: an hypothetical hydrophobic-like temperature dependence
of the model interaction strength $-\epsilon_0/k_{\rm B}T$ that drives 
folding kinetics (left vertical scale). Dashed line: an hypothetical 
temperature dependence of the intrinsic conformational transition rate
$1/A(T)$ relative to a reference rate $1/A_0$ (right vertical scale). More 
analytical details of this physical picture are provided in refs.~26 and 27. 
(Note that a typographical error should be corrected in the caption for 
Figure~4 of ref.~26: ``$\Delta S_0=5.2$'' should read ``$\Delta S_0=-5.2$.'')
(c) Upper curves: Temperature-dependent folding and unfolding rates 
obtained by combining the solid curve in (b) and data from Figure~5a
(equations~8 and 9, temperature-dependent intrinsic conformational 
transition rate
not taken into account). Lower curves: Temperature-dependent folding and 
unfolding rates obtained by combining the solid curve and dashed line in 
(b) and data from Figure~5a (equation~10, temperature-dependent intrinsic 
conformational transition rate taken into account). 
(d) Included for comparison are the temperature-dependent CI2 folding and 
unfolding rates at 25$^\circ$C and pH 6.3 from the experiments of Jackson 
and Fersht (adapted from Figure~4 of ref.~37). The upper and lower folding 
curves were for 0 M and 1.5 M GdnHCl respectively, the unfolding curve 
was measured at 0 M GdnHCl. 
\\

\noindent
{\bf Figure 8.} $\quad$
Schematics of hypothetical and proposed energy landscapes for protein 
folding. (a) A golf-course or ``Levinthal'' landscape. (b) A funnel 
landscape.  (c) A ``near-Levinthal'' scenario.
\\


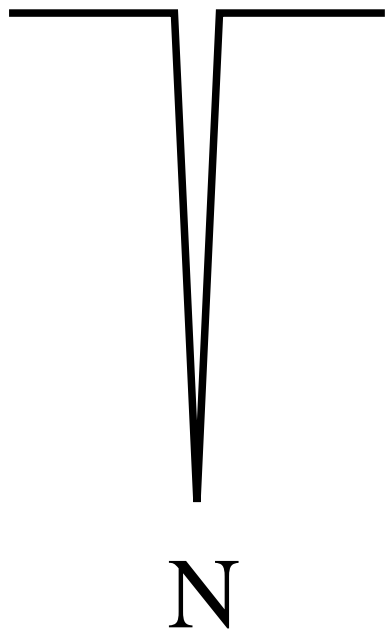
\end{document}